# Optomechanical system with tunable dissipative and dispersive couplings

Quansen Wang,[1] Yuefan Wu,[1] Doudou Wang,[1] Genyuan Xu,[1]
Jiawei Liang,[1] Qiang Zhang,[1,2,a)] and Yongmin Li[1,2,b)]

*[1]State Key Laboratory of Quantum Optics and Quantum Optics Devices, Institute of Opto-Electronics, Shanxi University, Taiyuan 030006, China*

*[2]Collaborative Innovation Center of Extreme Optics, Shanxi University, Taiyuan 030006, China*

We demonstrate an optomechanical system with tunable dissipative and dispersive couplings using a Fabry-Perot cavity and a string mechanical resonator. By varying the diameter and material of the mechanical resonator, and the relative location between the mechanical resonator and the cavity, the relative strengths of dissipative and dispersive coupling could be tuned continuously from dissipation-dominated regime to dispersion-dominated regime. In our experiments, the dissipative-to-dispersive coupling ratios of 1.3 and 0.6 are achieved by using two different mechanical resonators, corresponding to a transition from dissipation-dominated to dispersion-dominated optomechanical system. Theoretically, the coupling ratio could be tuned from 25 to 0.02 by optimizing the mechanical resonator, spanning over three orders of magnitude. These two distinct coupling regimes are achieved with the same experimental platform. The capability to freely adjust the coupling ratio provides a versatile platform for exploring quantum effects of massive mechanical resonators and quantum-limited measurements.

a) qzhang@sxu.edu.cn.
b) yongmin@sxu.edu.cn.

Cavity optomechanics, which studies the interaction between intracavity optical fields and mechanical resonators[1-3], provides a versatile platform for fundamental physics and sensing applications. For fundamental physics studies, cavity optomechanics enables the coherent control of quantum states of mechanical motion, extending the principles of quantum mechanics to the macroscopic scale[4-18]. For sensing applications, cavity optomechanics has enabled precision measurements of various physical quantities approaching or even surpassing the standard quantum limit[19-35]. The aforementioned optomechanical systems are primarily based on dispersive coupling. Compared with dispersive coupling, dissipative optomechanical systems exhibit unique advantages for ground-state cooling of massive mechanical resonators[36-39]. In addition, similarly to dispersive optomechanical systems, dissipative optomechanical systems can also achieve quantum-limited measurements of mechanical resonators[40]. Dissipative coupling has been experimentally observed in various cavity optomechanical systems, including whispering-gallery-mode cavities[41], photonic crystal cavities[42], and Fabry-Perot (FP) cavities[43-45]. However, ground state cooling of low-frequency mechanical resonators and quantum-limited measurements have not been realized in dissipative optomechanical systems. Significant intracavity losses and an inappropriate ratio between dissipative and dispersive coupling are the main obstacles for those systems[46,47]. Therefore, constructing a cavity optomechanical system with low intracavity loss and tunable dissipative-to-dispersive coupling ratio is essential for ground state cooling of low-frequency mechanical resonators and quantum-limited measurements.

In this letter, an optomechanical system with tunable dissipative and dispersive couplings is proposed and demonstrated using a FP cavity and a string mechanical resonator. By adjusting the parameters of the mechanical resonator and the relative location between the mechanical resonator and the FP cavity, the relative strength of dissipative and dispersive couplings could be tuned about three orders of magnitude from 25 to 0.02 in theory. In experiments, a 10 μm-diameter iron-wire resonator and a 5 μm-diameter fiber-optic resonator are used to couple with the FP cavity with a finesse of 13000, respectively. The corresponding maximum and minimum values of relative strength of dissipative and dispersive couplings are 1.3 and 0.6, validating the theoretical predictions. Our work offers a promising optomechanical system for exploring quantum effects of massive mechanical resonators and quantum-limited measurements.

The optomechanical system comprises a symmetric concave-mirror FP cavity and a slender cylinder mechanical resonator as shown in Fig. 1(a). As the mechanical resonator is inserted progressively deeper into the cavity, the spatial overlap region between the mechanical resonator and the intracavity standing wave field increases effectively. Besides the fixed decay rate $\gamma_1$ of the FP cavity itself, the mechanical resonator scatters a part of the intracavity light out of the cavity, and introduces an additional position-dependent dissipative decay rate $\gamma_2$ ($x$). As the position of the mechanical resonator varies, the effective transmissivity and reflectivity of the mechanical resonator are dynamically modulated, resulting in a shift in the cavity resonant frequency $\omega_c$ (dispersive effect) and a broadening of the linewidth (dissipative effect). The transfer matrix $M$ of the optomechanical system can be expressed as follows[48]:

$$M = M_{\mathrm{L}} M_{-2/\mathrm{L}} M_{\mathrm{m}} M_{2/\mathrm{L}} M_{\mathrm{R}}, \tag{1}$$

where $M_L$ and $M_R$ are the transfer matrices of the left and right cavity mirrors, $M_{-2/L}$ and $M_{2/L}$ are the transfer matrices of the cavity sections on either side of the mechanical resonator, and $M_m$ is the transfer matrix of the mechanical resonator.The definitions and derivations of the individual submatrices are given in the Supplementary Material. Accordingly, the reflectance $C_R$, transmittance $C_T$, and scattering rate $C_S$ of the optomechanical system are given by:

$$C_{\mathrm{R}} = \left|\frac{-M_{21}}{M_{22}}\right| \qquad C_{\mathrm{T}} = \left|\frac{M_{11}M_{22} - M_{12}M_{21}}{M_{22}}\right| \qquad C_{\mathrm{S}} = 1 - C_{\mathrm{R}} - C_{\mathrm{T}}. \tag{2}$$

Using these expressions, we theoretically calculated the reflection, transmission, and scatter patterns of the optomechanical system when an iron-wire mechanical resonator (diameter $D_M$ = 10 μm) is moved from the outside to the inside of the optical cavity, the results are shown in Figs. 1(b)–(d). Fig. 1(b) shows that as the mechanical resonator is inserted deeper into the cavity, the optical cavity's resonant frequency increases and its reflectance decreases. From Figs. 1(b) and 1(c), the sum of transmittance and reflectance is less than unity, indicating that the mechanical resonator scatters part of the light out of the cavity. This conclusion is further confirmed by the scattering rate shown in Fig. 1(d).The laser wavelength $\lambda$ is 1064 nm, the cavity length $L$ is 17 cm, the waist radius $\omega$ is 80 μm and the reflectivities of the mirrors are $R_a = R_b$ = 99.98%. From Figs. 1(b)-(d), two key coupling parameters of the intracavity mode—the dissipative coupling constants $g_\gamma = d\gamma_2/dx$ and the dispersive coupling constants $g_\omega = d\omega_c/dx$, which together determine the response characteristics of the system in optomechanical interaction, can be determined.

The dispersive and dissipative coupling constants depend on the diameter and material of the mechanical resonator and the relative location between the mechanical resonator and the FP cavity. Figs. 1(e) and (f) show the simulated dispersion and dissipation induced by two fiber-optic mechanical resonators with different diameters as they are inserted into the cavity along the beam waist position. The red solid curve, red dashed curve, and purple curve represent the dispersive coupling constant, dissipative coupling constant, and dissipative-dispersive coupling ratio, respectively. For $D_M$ = 1.29 μm, dissipation exceeds dispersion by a ratio of approximately 25, while for $D_M$ = 2.28 μm, the corresponding ratio is approximately 0.02. Therefore, we can achieve a transition with the same system from dissipation-dominated (dissipative rate greater than dispersive rate) to dispersion-dominated (dispersive rate greater than dissipative rate) regimes by tuning the parameters of the mechanical resonator. This ability to tune the dissipation-dispersion ratio freely allows our system to meet the specific coupling requirements of different precision-measurement scenarios flexibly.

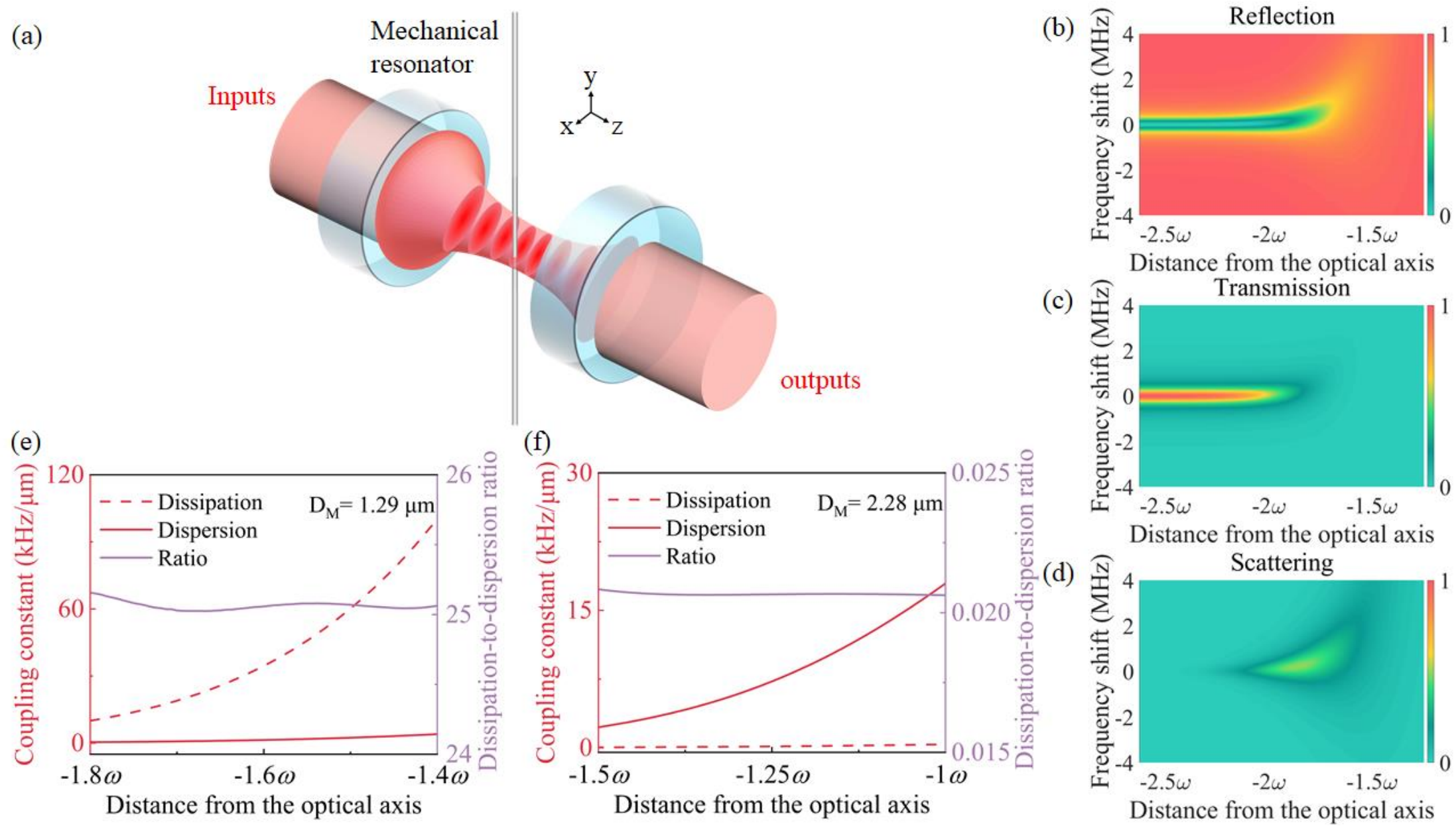

FIG. 1. Optomechanical system with tunable dispersive and dissipative coupling. (a) Schematic diagram of the optomechanical system. (b)-(d) Reflection, transmission, and scattering intensity signal of the optomechanical system when the mechanical resonator is positioned at different radial locations across the optical field cross-section. Frequency shift denotes the resonant-frequency change of the optical FP cavity induced by the mechanical resonator. (e)-(f) Dissipation-dominated and dispersion-dominated optomechanical systems using different mechanical resonators.

To validate the aforementioned phenomena, an optomechanical system is built based on two FP cavities and a mechanical resonator. The mechanical resonator is mounted on a high-precision motorized stage with step sizes on the order of tens of nanometers to adjust its relative position in one of the FP cavities, as depicted in Fig. 2(a). A dual-cavity structure is designed to effectively suppress the influence of environmental disturbances, such as the temperature drift and mechanical vibrations. The cavity containing the mechanical resonator is defined as the experimental cavity, while the other cavity without the mechanical resonator serves as the reference cavity. Compared with the single-cavity configuration, the reference cavity and the experimental cavity are mounted on the same Invar base and share the same piezoelectric actuation system. Therefore, systematic errors arising from the Invar base and the actuation system affect both cavities identically. By subtracting the output signals of the two cavities, these errors can be effectively eliminated, and the dependence relationship of the resonant frequency and linewidth of the experimental cavity on the position of the mechanical oscillator are obtained. This approach significantly improves the measurement accuracy.

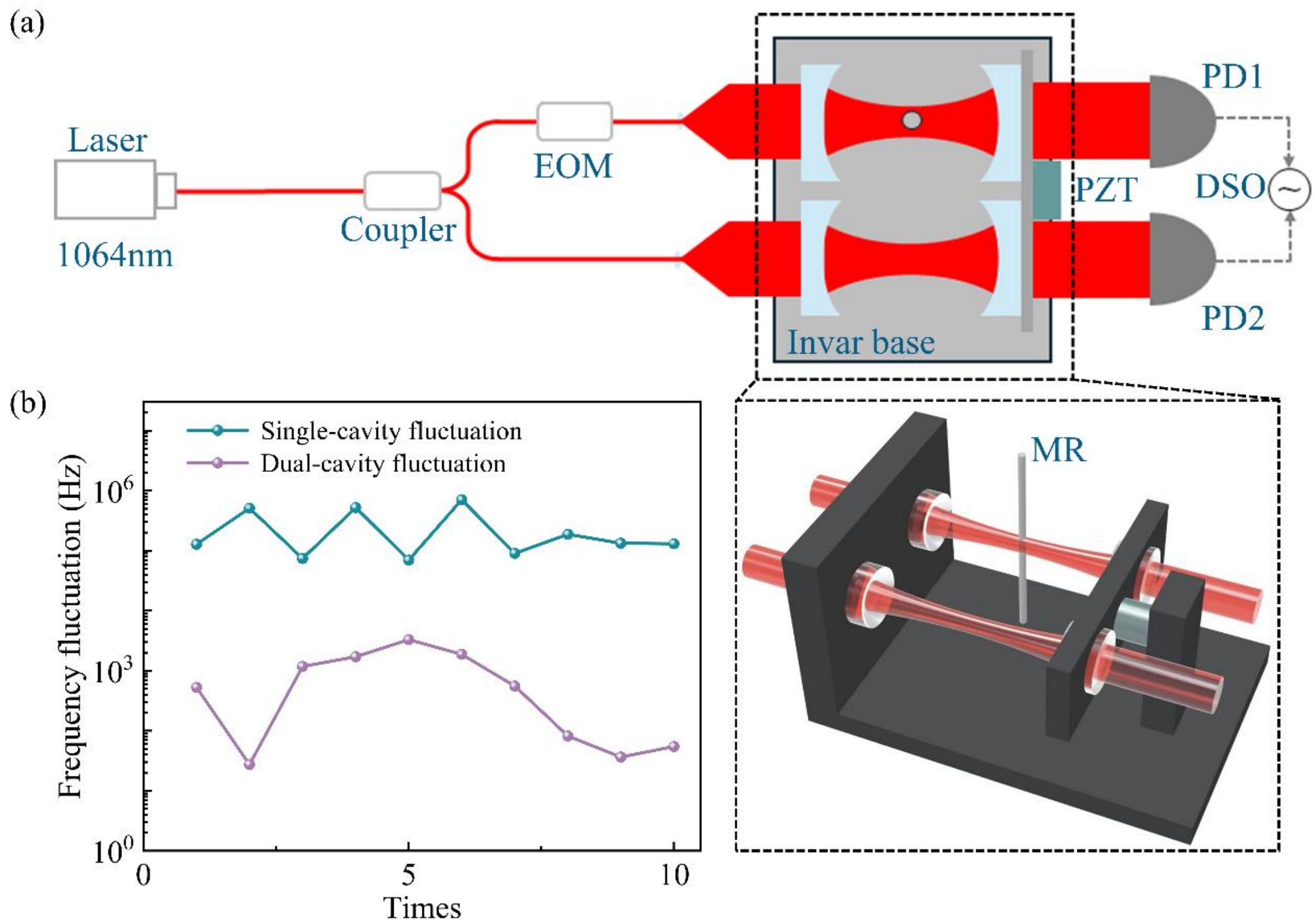

FIG. 2. Schematic description of the experimental setup. (a) Simplified diagram of the setup; the inset shows the detailed cavity structure. EOM: Electro-optic modulator; PZT: Piezoelectric Transducers; PD: Photodetector; DSO: Digital storage oscilloscope; MR: mechanical resonator. (b) Single-cavity and dual-cavity frequency drifts obtained from multiple sets of measurements.

As shown in Fig. 2(b), the resonant frequency of the single-cavity system exhibits typical fluctuations on the order of $10^5$ Hz due to environmental disturbances. In contrast, with the common-substrate and common-drive dual-cavity structure, these fluctuations are suppressed effectively to the order of $10^3$ Hz, a reduction of over two orders of magnitude. Owing to the enhanced performance, the cavity frequency shifts and linewidth broadenings induced by mechanical resonator displacement are resolved over extended timescales, thereby providing a stable platform for optomechanical interaction with tuning the dissipative-dispersive coupling ratio.

Using the dual-cavity optomechanical system, we investigate how the position of the mechanical resonator affects the cavity transmission spectrum. First, the changes of the linewidth and the resonant frequency are measured as mechanical resonators of different materials and diameters are placed at different positions in the optical cavity. Figs. 3(a) and 3(b) show the variation trends of the linewidth and resonant frequency shift for the 10 μm iron-wire and the 5 μm fiber-optic mechanical resonator, respectively. The blue and red points represent the experimentally measured linewidth and frequency shift, and the standard deviation is presented as the error bars; the blue dashed lines and red solid lines represent the theoretically predicted linewidth and frequency shift, respectively. Experimentally, the mechanical resonator is moved to two different positions. The resonant frequency and linewidth are recorded at each position, and the changes between them are calculated. The variations of the resonant frequency and linewidth per displacement are then assigned as the dispersive and dissipative coupling values at the midpoint between these positions. For the iron-wire mechanical resonator, both the linewidth and the resonant frequency shift exhibit a monotonically increasing trend as it moves to the optical axis of the cavity, as shown in Fig. 3(a). In contrast, when the fiber-optic mechanical resonator is used, the finer diameter results in a smaller spatial overlap with the optical field and a relatively limited perturbation to the cavity, as shown in Fig. 3(b). Consequently, although both dissipative and dispersive coupling also increase as the fiber-optic mechanical resonator moves towards the optical axis of the cavity, the changes of the linewidth and the resonant frequency are significantly slower than those of the iron-wire mechanical resonator.

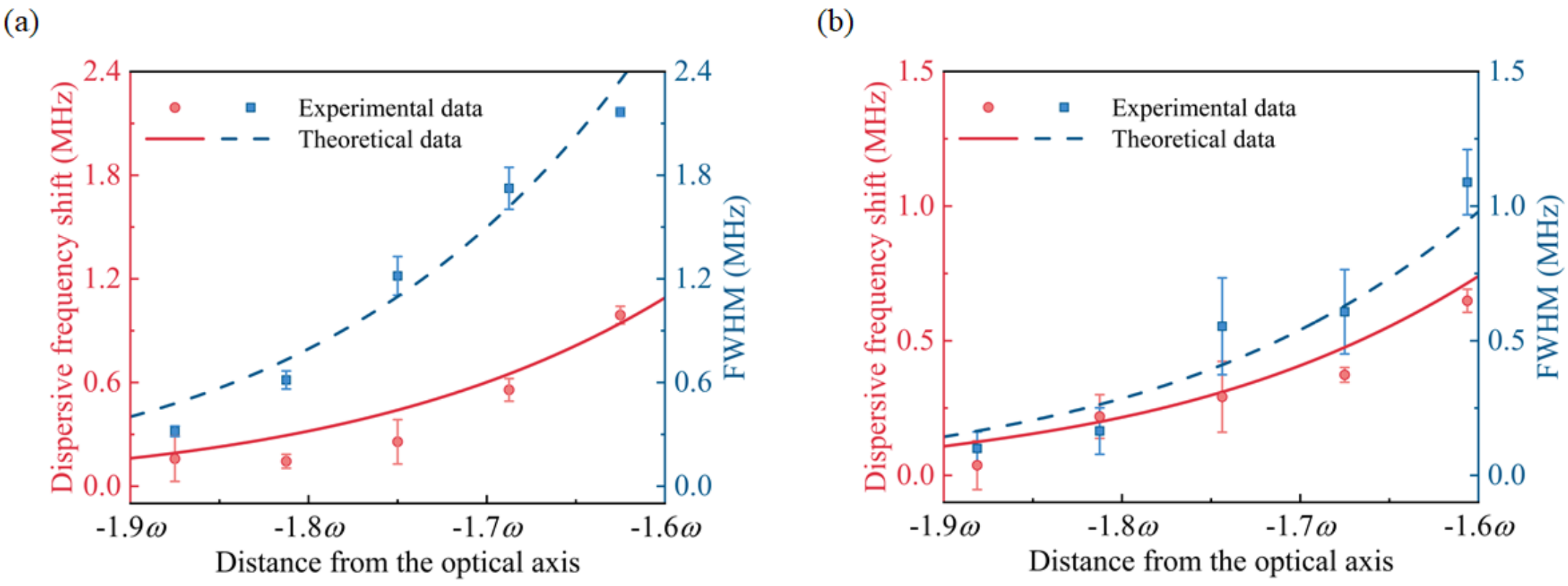

FIG. 3. Dispersive frequency shift of the cavity mode resonant frequency and the linewidth of cavity as the mechanical resonator is gradually moved towards the optical axis of the cavity. The experimental and calculated values of the linewidth and resonant frequency shift are shown for the 10 μm iron-wire (a) and the 5 μm fiber-optic (b) mechanical resonator, respectively.

From Fig. 4, the dissipative and dispersive coupling strengths are determined. Fig. 4 shows the corresponding dissipative and dispersive coupling strengths for the 10 μm iron-wire and the 5 μm fiber-optic mechanical resonator, respectively. The blue and red points represent the experimentally measured dissipative and dispersive coupling strengths, and the blue dashed lines and red solid lines represent the theoretical dissipative and dispersive coupling strengths. For the iron-wire mechanical resonator, both of the dissipative and dispersive coupling strengths gradually increase as the mechanical resonator moves towards the optical axis of the cavity. Notice that the dissipative coupling strength is larger than the dispersive coupling strength. The system thus enters a dissipation-dominated regime, with a dissipative-to-dispersive coupling ratio of 1.3, as shown in Fig. 4(a). For the fiber-optic mechanical resonator, although both dissipative and dispersive coupling strengths also increase, the system enters a dispersion-dominated regime with a dissipative-to-dispersive coupling ratio of 0.6, as shown in Fig. 4(b).

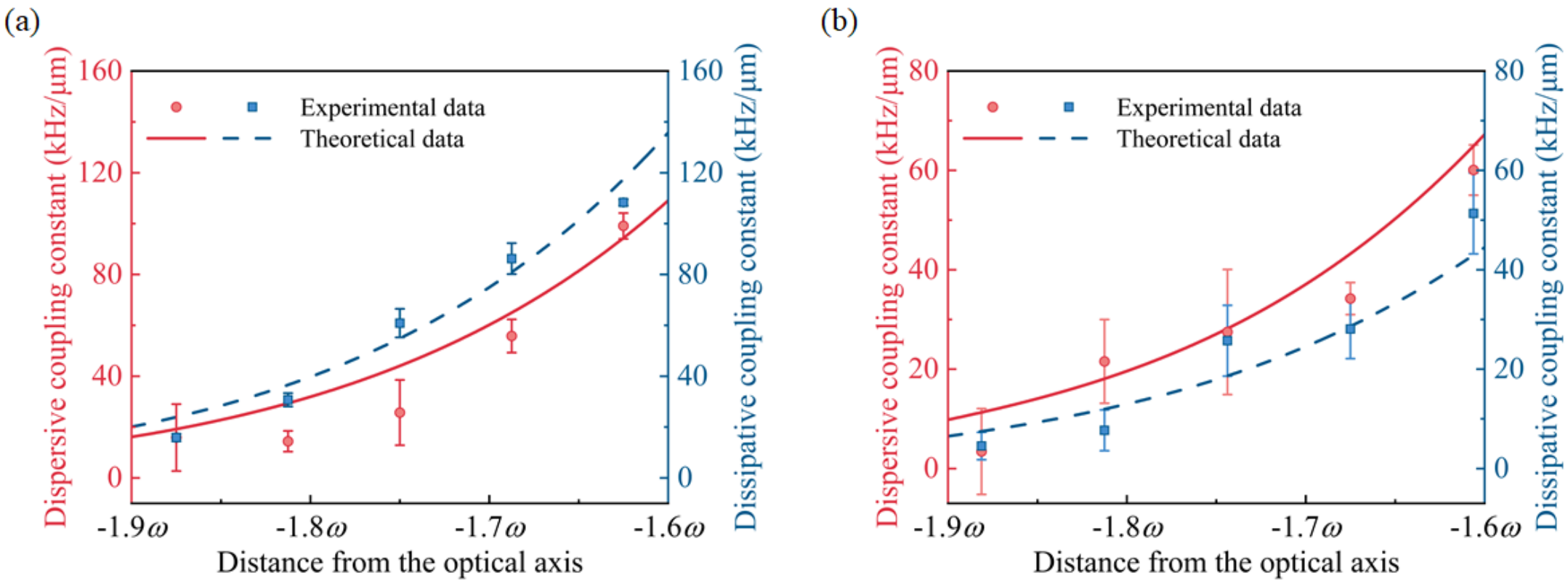

FIG. 4. Dissipative and dispersive coupling strengths as the mechanical resonator is gradually moved towards the optical axis of the cavity. The experimental and calculated values of the dissipative and dispersive coupling constants are presented for the 10 μm iron-wire (a) and the 5 μm fiber-optic (b) mechanical resonator, respectively.

The results presented in Figs. 3 and 4 clearly show that we have achieved tuning of dissipative and dispersive coupling strengths, laying the foundation for applications such as ground state cooling and quantum-limited measurements. The difference between the theoretical values and the experimental data results from the measuring errors of the dissipative and dispersive coupling strengths in our experiments. As shown in Fig. 2, although the common-substrate and common-drive dual-cavity structure reduces the impact of environmental disturbances effectively, the residual fluctuations still remain. This hinders the measurement of smaller dissipative and dispersive coupling and observe large relative strength of dissipative and dispersive couplings (25 to 0.02 as shown in Fig. 1(e) and (f). As a next step, we plan to place the system in a vacuum environment and stabilize the temperature of the experimental setup to further suppress the influence of thermal fluctuations and airflow on the optomechanical system. Meanwhile, as predicted by the theory of dissipative and dispersive hybrid optomechanical coupling[40,47], when the proposed system operates at cryogenic temperatures and uses a high-Q fiber mechanical resonator[49], ground-state cooling and quantum-limited measurements of the mechanical resonator in the unresolved-sideband regime should be feasible.

In summary, we have built an optomechanical system consisting of dual-Fabry-Perot cavities and a movable string mechanical resonator. By adjusting the position, diameter, and material of the mechanical resonator, the optomechanical system could be tuned flexibly from dissipation-dominated to dispersion-dominated regimes. Using a common-substrate and common-drive referencing optical cavity, environmental perturbations are suppressed effectively by about two orders of magnitude. The experimental results agrees with the theoretical predictions, confirming the tunable dissipation and dispersion couplings of the proposed optomechanical system. This versatile system offers a promising route for ground state cooling of massive mechanical resonators and quantum-limited measurements.

## SUPPLEMENTARY MATERIAL

See the supplementary material for the additional details on a theoretical model for the proposed optomechanical system with tunable dissipative and dispersive couplings and the resonant frequencies and $Q$ factors of the mechanical resonators.

This work was supported by the National Natural Science Foundation of China (grants no.92565108 and 12174232), Innovation Program for Quantum Science and Technology (grants no. 2023ZD0300400), and Basic Research Program of Shanxi Province (grants no. 202503021211060).

## AUTHOR DECLARATIONS

### Conflict of Interest

The authors have no conflicts to disclose.

### Author Contributions

**Quansen Wang:** Data curation (equal); Formal analysis (equal); Investigation (equal); Methodology (equal); Soft-ware (equal); Visualization (equal); Writing - original draft (equal). **Yuefan Wu:** Formal analysis (equal); Investigation (equal); Soft-ware (equal); Visualization (equal). **Doudou Wang:** Formal analysis (equal); Methodology (equal). **Genyuan Xu:** Formal analysis (equal); Soft-ware (equal). **Jiawei Liang:** Visualization (equal). **Qiang Zhang:** Conceptualization (equal); Funding acquisition (equal); Project administration (equal); Resources (equal); Supervision(equal); Writing - review & editing (equal). **Yongmin Li:** Funding acquisition (equal); Project administration (equal); Resources (equal); Supervision(equal); Writing - review & editing (equal).

## DATA AVAILABILITY

The data that support the findings of this study are available from the corresponding author upon reasonable request.

# Supplementary Information for:

# Optomechanical system with tunable dissipative and dispersive couplings

Quansen Wang,[1] Yuefan Wu,[1] Doudou Wang,[1] Genyuan Xu,[1] Jiawei Liang,[1] Qiang Zhang,[1,2,a)] and Yongmin Li[1,2,b)]

[1]State Key Laboratory of Quantum Optics and Quantum Optics Devices, Institute of Opto-Electronics, Shanxi University, Taiyuan 030006, China

[2]Collaborative Innovation Center of Extreme Optics, Shanxi University, Taiyuan 030006, China

CONTENTS

## SI1 Transfer matrix and scattering matrix

The evolution of optical fields in an optical system is commonly described in matrix form. Two widely used representations are the ABCD-type transfer matrix $M$ and the scattering matrix $S$, which can be converted into each other.

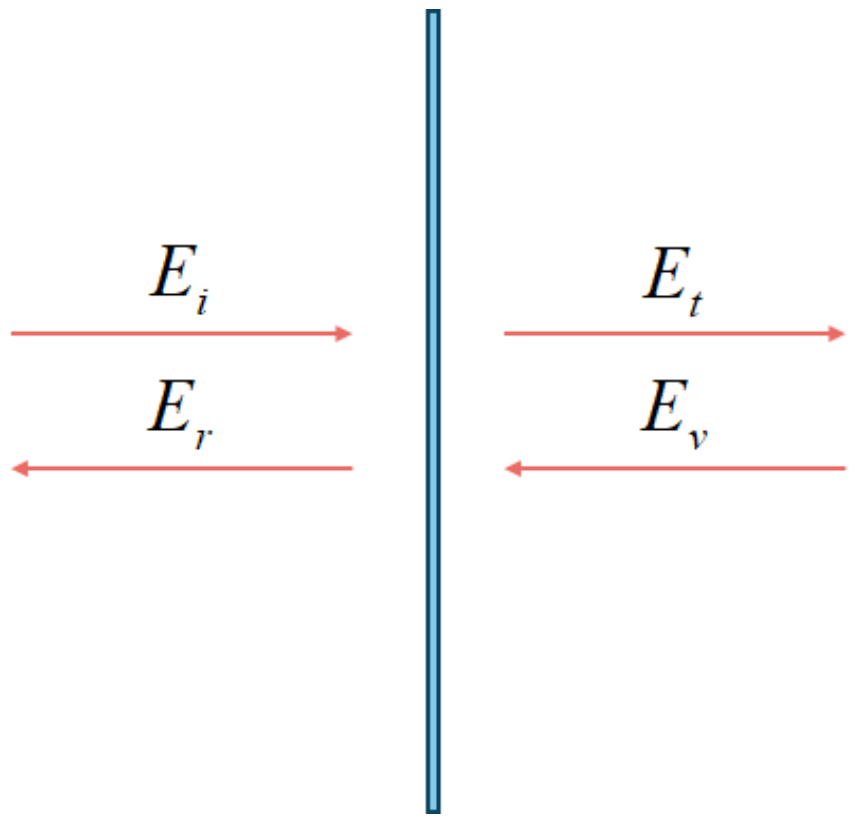

FIG. S1. Schematic physical model of plane-wave propagation through a dielectric **medium**.

Alt text:
FIG. S1: [A plane wave incident from the left undergoes reflection and transmission at the dielectric **medium**]

As shown in Fig. S1, a plane wave is incident from the left and undergoes reflection and transmission at the dielectric **medium**. The output fields $E_r$ and $E_t$ can be expressed as functions of the input fields $E_i$ and $E_v$[1]:

$$\begin{pmatrix} E_t \\ E_r \end{pmatrix} = S \begin{pmatrix} E_i \\ E_v \end{pmatrix} = \begin{pmatrix} S_{11} & S_{12} \\ S_{21} & S_{22} \end{pmatrix} \begin{pmatrix} E_i \\ E_v \end{pmatrix} \quad S = \begin{pmatrix} t_{12} & r_{21} \\ r_{12} & t_{21} \end{pmatrix}, \tag{S1}$$

where $t_{ij}$ and $r_{ij}$ are obtained from the Fresnel formulas:

$$t_{ij} = \frac{2n_i}{n_i + n_j} \quad r_{ij} = \frac{n_i - n_j}{n_i + n_j}. \tag{S2}$$

The scattering-matrix formalism is typically used to describe the relation between incident and outgoing waves. However, for multilayer structures, it is often more convenient to use a transfer matrix to describe the entire system:

$$\begin{pmatrix} E_t \\ E_v \end{pmatrix} = M \begin{pmatrix} E_i \\ E_r \end{pmatrix} = \begin{pmatrix} M_{11} & M_{12} \\ M_{21} & M_{22} \end{pmatrix} \begin{pmatrix} E_i \\ E_r \end{pmatrix}, \quad M = \frac{1}{t_{21}} \begin{pmatrix} 1 & r_{21} \\ -r_{12} & 1 \end{pmatrix}, \tag{S3}$$

The transfer matrix relates the waves on one side of the system to those on the other side. The two matrix representations can be converted into each other. Writing Eq. (S1) as a set of equations gives:

$$\begin{cases} E_t = S_{11} E_i + S_{12} E_v \\ E_r = S_{21} E_i + S_{22} E_v \end{cases}. \tag{S4}$$

After algebraic rearrangement, the transfer-matrix form relating $E_t$, $E_v$, $E_i$, and $E_r$ can

be obtained as:

$$\begin{aligned} E_t &= \left( S_{11} - \frac{S_{12}S_{21}}{S_{22}} \right) E_i + \frac{S_{12}}{S_{22}} E_r \\ E_v &= -\frac{S_{21}}{S_{22}} E_i + \frac{1}{S_{22}} E_r \end{aligned} \Rightarrow \begin{pmatrix} E_t \\ E_v \end{pmatrix} = \frac{1}{S_{22}} \begin{pmatrix} S_{11}S_{22} - S_{12}S_{21} & S_{12} \\ -S_{21} & 1 \end{pmatrix} \begin{pmatrix} E_i \\ E_r \end{pmatrix}. \quad \text{(S5)}$$

## SI2 Reflection and transmission coefficients of a dielectric medium

We consider the reflection and transmission coefficients of a plane wave normally incident on a homogeneous dielectric **medium** with refractive index $n$ and thickness $L_m$ (Fig. S2). The output and input fields can be written as:

$$\begin{pmatrix} E_t \\ E_v \end{pmatrix} = M \begin{pmatrix} E_i \\ E_r \end{pmatrix} \qquad M = M_1 M_{L_m} M_2, \tag{S6}$$

where

$$\begin{aligned} M_1 &= \frac{1}{t_1}\begin{pmatrix} 1 & -r_1 \\ -r_1 & 1 \end{pmatrix} \quad M_2 = \frac{1}{t_2}\begin{pmatrix} 1 & r_2 \\ r_2 & 1 \end{pmatrix} \\ M_{L_m} &= \begin{pmatrix} e^{iknL_m} & 0_1 \\ 0 & e^{-iknL_m} \end{pmatrix}, \end{aligned} \tag{S7}$$

The reflection and transmission coefficients are then obtained as follows, where $\varphi = nkL$, $k=2\pi/\lambda$ is the wave vector in vacuum:

$$\begin{aligned} C_r &= \left| \frac{-M_{21}}{M_{22}} \right| = \frac{(1-n^2)\sin\varphi}{2in\cos\varphi + (1+n^2)\sin\varphi} \\ C_t &= \left| \frac{1}{M_{22}} \right| = \frac{2ni}{2in\cos\varphi + (1+n^2)\sin\varphi}. \end{aligned} \tag{S8}$$

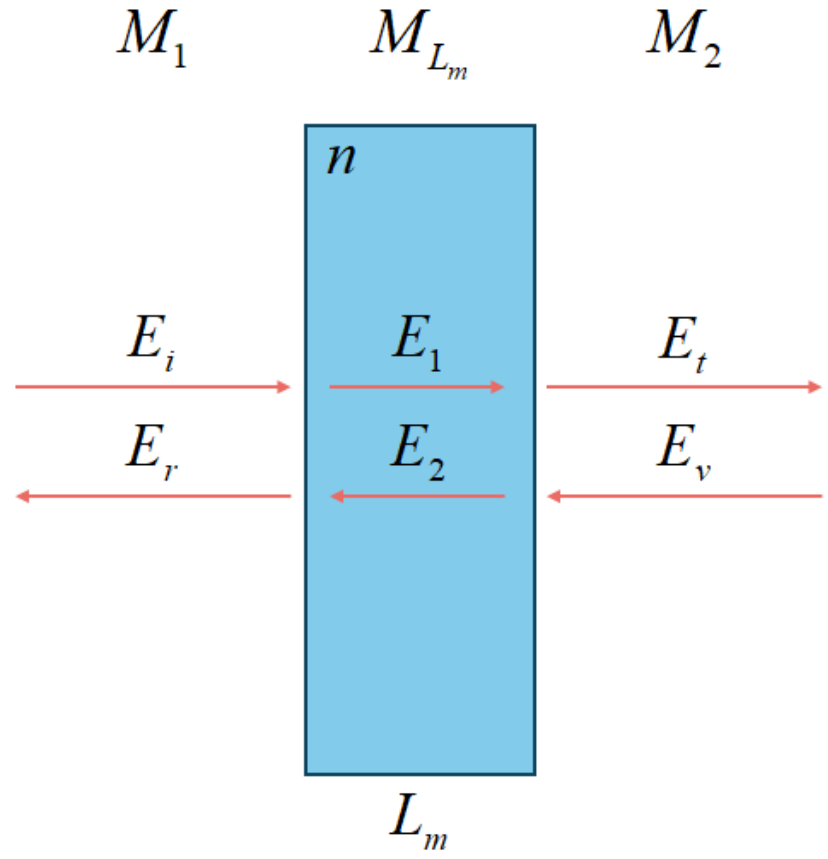

FIG. S2. Schematic electric-field distribution in a dielectric **medium** with refractive index $n$ and thickness $L_m$.

Alt text:
FIG. S2: [A plane wave incident from the left undergoes reflection and transmission at the dielectric **medium** with refractive index $n$ and thickness $L_m$.]

## SI3 Fabry-Perot cavity containing a dielectric medium

To facilitate the analysis of the behavior of a simply supported beam mechanical resonator inside a Fabry-Perot (FP) cavity, we first examine the characteristics of a dielectric medium placed in the cavity.

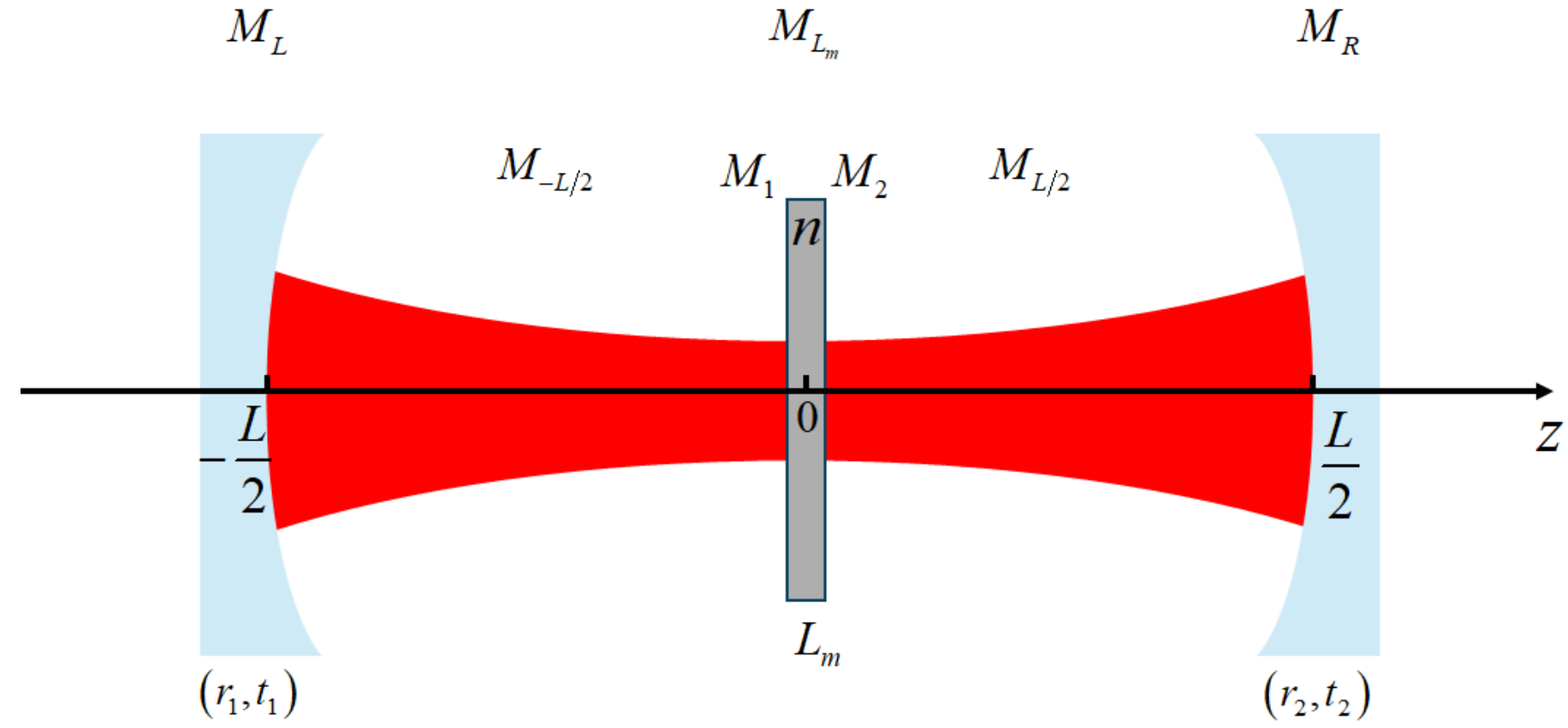

FIG. S3. Schematic of a dielectric medium in a Fabry-Perot cavity.

Alt text:
FIG. S3: [A double-concave FP cavity of length $L$ contains a dielectric medium of thickness $L_m$ and refractive index $n$.]

As shown in Fig. S3, a dielectric medium with thickness $L_m$ and refractive index $n$ is inserted into a double-concave FP cavity of length $L$. The reflection and transmission coefficients of the left and right cavity mirrors are denoted by $r_1$, $t_1$, $r_2$, and $t_2$, respectively. We assume that the dielectric medium is non-absorbing. Under this assumption, the dielectric medium divides the FP cavity into three regions: the left sub-cavity, extending from the input mirror to the left surface of the dielectric medium, with length $L_1 = L/2 - L_m/2$; the middle region, corresponding to the dielectric medium itself; and the right sub-cavity, extending from the right surface of the dielectric medium to the output mirror, with length $L_2 = L/2 - L_m/2$.

We use the characteristic transfer-matrix formalism to calculate the cavity reflection and transmission. Owing to the presence of the dielectric medium, the overall transfer matrix $M$ of the cavity is decomposed into the following components: the transfer matrices of the left and right cavity mirrors, $M_L$ and $M_R$; the propagation matrices of the Gaussian beam in the two sub-cavities, $M_{-L/2}$ and $M_{L/2}$; the interface transfer matrices of the two surfaces of the dielectric medium, $M_1$ and $M_2$; and the propagation matrix inside the dielectric medium $M_{L_m}$. Thus,

$$M = M_L M_{-L/2} M_1 M_{L_m} M_2 M_{L/2} M_R, \tag{S9}$$

where

$$M_L = \frac{1}{t_1}\begin{pmatrix} 1 & -r_1 \\ -r_1 & 1 \end{pmatrix} \quad M_R = \frac{1}{t_2}\begin{pmatrix} 1 & r_2 \\ r_2 & 1 \end{pmatrix},$$
$$M_1 = \frac{1}{t_m}\begin{pmatrix} 1 & -r_m \\ -r_m & 1 \end{pmatrix} \quad M_2 = \frac{1}{t_m}\begin{pmatrix} 1 & r_m \\ r_m & 1 \end{pmatrix}, \tag{S10}$$
$$M_{-L/2} = \begin{pmatrix} e^{ikL_1} & 0 \\ 0 & e^{-ikL_1} \end{pmatrix} \quad M_{L/2} = \begin{pmatrix} e^{ikL_2} & 0 \\ 0 & e^{-ikL_2} \end{pmatrix} \quad M_{L_m} = \begin{pmatrix} e^{iknL_m} & 0 \\ 0 & e^{-iknL_m} \end{pmatrix},$$

with $r_m = (1 - n)/(1 + n)$, $t_m = 2/(1 + n)$, and $\tilde{t}_m = 2n/(1 + n)$.

The reflection and transmission coefficients are given by:

$$C_r = \left| -\frac{M_{21}}{M_{22}} \right|, \mathrm{C}_t = \left| \frac{1}{M_{22}} \right|. \tag{S11}$$

### SI4 Fabry-Perot cavity with a partially inserted beam mechanical resonator

We first assume that only the fundamental mode exists in the cavity. The mechanical resonator (MR) is modeled as having a square cross section with dimensions $L_x$ and $L_z$, and is assumed to be infinitely long along the $y$ direction. To obtain the transfer matrix of the MR inserted into the Gaussian mode, we treat the interaction as a shadow effect: the mechanical resonator acts as a small obstacle that partially blocks the incident Gaussian beam and thereby produces a shadowed region in the optical field. This approximation is used to describe the perturbation of the optical field induced by the resonator. The center position of the resonator is denoted by $\boldsymbol{r}_0 = (x_0, y_0, z_0)$, as shown in Fig. S4.

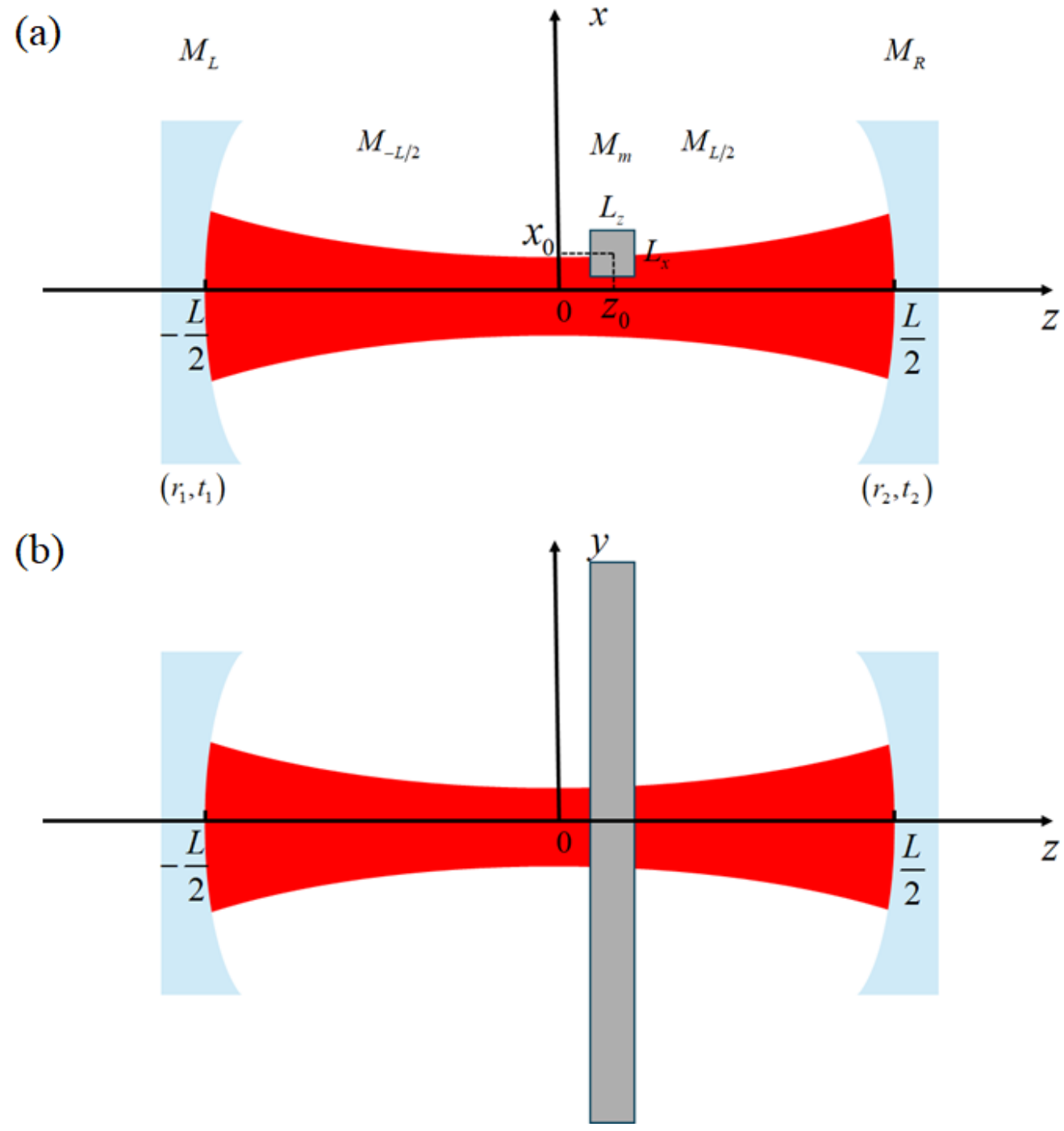

FIG. S4. Schematic of the system model containing a beam resonator. (a) Schematic in the $x$-$z$ plane; (b) schematic in the $y$-$z$ plane.

Alt text:
FIG. S4: [A partially inserted mechanical resonator is placed into a double-concave FP cavity of length $L$. (a) Along the x-z direction, the resonator is partially inserted into the optical field. (b) Along the y-z direction, the resonator completely passes through the optical field.]

We assume that light propagates as a plane wave inside the mechanical resonator and neglect the divergence of the Gaussian beam. Under these assumptions, the transmission coefficient $t_M$ and reflection coefficient $r_M$ of the mechanical resonator depend on the coordinate $\boldsymbol{r}$:

$$r_M = \begin{cases} C_r & \mathbf{r} \in \text{MR} \\ 0 & \mathbf{r} \notin \text{MR} \end{cases}$$
$$t_M = \begin{cases} C_t & \mathbf{r} \in \text{MR} \\ e^{ikL_z} & \mathbf{r} \notin \text{MR} \end{cases} . \tag{S12}$$

$C_r$ and $C_t$ are defined as in Eq. (S8). The effective reflection and transmission coefficients of the MR can then be obtained through a projection operation:

$$r_{MR} = \iint d\mathbf{r}\, r_M(\mathbf{r}) \rho^2\left(\mathbf{r}, z_0 - \frac{L_z}{2}\right)$$
$$t_{MR} = \iint d\mathbf{r}\, t_M(\mathbf{r}) \rho^2\left(\mathbf{r}, z_0 + \frac{L_z}{2}\right), \tag{S13}$$

where $\rho$ is the amplitude of the Gaussian field as given by Eq. (S14), with $\omega(z)$ representing the beam radius:

$$\rho(\mathbf{r}, z) = A_0 \frac{\omega_0}{\omega(z)} e^{-\frac{\mathbf{r}^2}{\omega^2(z)}}, \tag{S14}$$

By using the error function and solving Eq. (S13), we obtain:

$$t_{MR} = e^{ikL_z} + \frac{1}{4}\left(C_t - e^{ikL_z}\right)\left[1 - erf\left(\frac{\sqrt{2}}{\omega(z)} y\right)\right]$$
$$\left[erf\left(\frac{\sqrt{2}}{\omega(z)}\left(x + \frac{L_x}{2}\right)\right) - erf\left(\frac{\sqrt{2}}{\omega(z)}\left(x - \frac{L_x}{2}\right)\right)\right], \tag{S15}$$

$$r_{MR} = \frac{C_r}{4}\left[1 - erf\left(\frac{\sqrt{2}}{\omega(z)} y\right)\right]$$
$$\left[erf\left(\frac{\sqrt{2}}{\omega(z)}\left(x + \frac{L_x}{2}\right)\right) - erf\left(\frac{\sqrt{2}}{\omega(z)}\left(x - \frac{L_x}{2}\right)\right)\right], \tag{S16}$$

$$erf(x) = \frac{2}{\sqrt{\pi}} \int_0^x e^{-t^2} dt. \tag{S17}$$

Here, we approximate the beam waists on the two sides of the mechanical resonator as $\omega(z_0 + L_z/2) \approx \omega(z_0 - L_z/2) \approx \omega(z_0)$.

In the experiments, the waist radius of the optical cavity is much smaller than the length of the mechanical resonator along the y direction. Therefore, when calculating the effective transmission coefficient $t_{MR}$ and reflection coefficient $r_{MR}$ of the mechanical resonator, y is treated as infinite.

Combining the results of Sections SI1-SI3, the transfer matrix of the system can be written as:

$$M = M_L M_{-2/L} M_m M_{2/L} M_R, \tag{S18}$$

where the individual transfer matrices are defined as:

$$M_L = \frac{1}{t_1}\begin{pmatrix} 1 & -r_1 \\ -r_1 & 1 \end{pmatrix} \quad M_R = \frac{1}{t_2}\begin{pmatrix} 1 & r_2 \\ r_2 & 1 \end{pmatrix}$$
$$M_{-L/2} = \begin{pmatrix} e^{i\left[\varphi_0(z_0)-\varphi_0(-L/2)\right]} & 0 \\ 0 & e^{-i\left[\varphi_0(z_0)-\varphi_0(-L/2)\right]} \end{pmatrix}$$
$$M_{L/2} = \begin{pmatrix} e^{i\left[\varphi_0(L/2)-\varphi_0(z_0)\right]} & 0 \\ 0 & e^{-i\left[\varphi_0(L/2)-\varphi_0(z_0)\right]} \end{pmatrix} \tag{S19}$$
$$M_m = \frac{1}{t_M}\begin{pmatrix} t_{MR}^2 - r_{MR}^2 & r_M \\ -r_M & 1 \end{pmatrix},$$

where the field phase is $\varphi_0(z) = kz - \psi(z)$, $\psi(z)$ is the Gouy phase, and $Z_R$ is the Rayleigh range, given by:

$$\Phi(z) = \arctan\left(\frac{z}{z_R}\right) \qquad z_R = \frac{\pi\omega_0^2}{\lambda}. \tag{S20}$$

Finally, the reflection, transmission, and scattering coefficients of the system can be expressed as:

$$C_R = \left|\frac{-M_{21}}{M_{22}}\right| \qquad C_T = \left|\frac{M_{11}M_{22} - M_{12}M_{21}}{M_{22}}\right| \qquad C_S = 1 - C_R - C_T. \tag{S21}$$

## SI5 Eigenfrequencies and $Q$ factors of the mechanical resonators

We measured the eigenfrequencies and $Q$ factors of the iron-wire mechanical resonator. Fig. S5 presents the normalized mechanical power spectra of the first-order and second-order modes, respectively, for the iron-wire mechanical resonator with a length of 5 mm and a diameter of 10 μm. The black dots represent the experimentally measured data, the red solid curves are Lorentzian fits to the data, and the insets show the mode shapes of the two resonant modes. For the first-order mode, the resonant frequency is $f_1$ = 1902 Hz, the linewidth is $\Gamma_1$ = 28.7 Hz, and the $Q$ factor is $Q_1 = f_1/\Gamma_1$ = 66; for the second-order mode, the resonant frequency is $f_2$ = 5232 Hz, the linewidth is $\Gamma_2$ = 42.9 Hz, and the $Q$ factor is $Q_2 = f_2/\Gamma_2$ = 122.

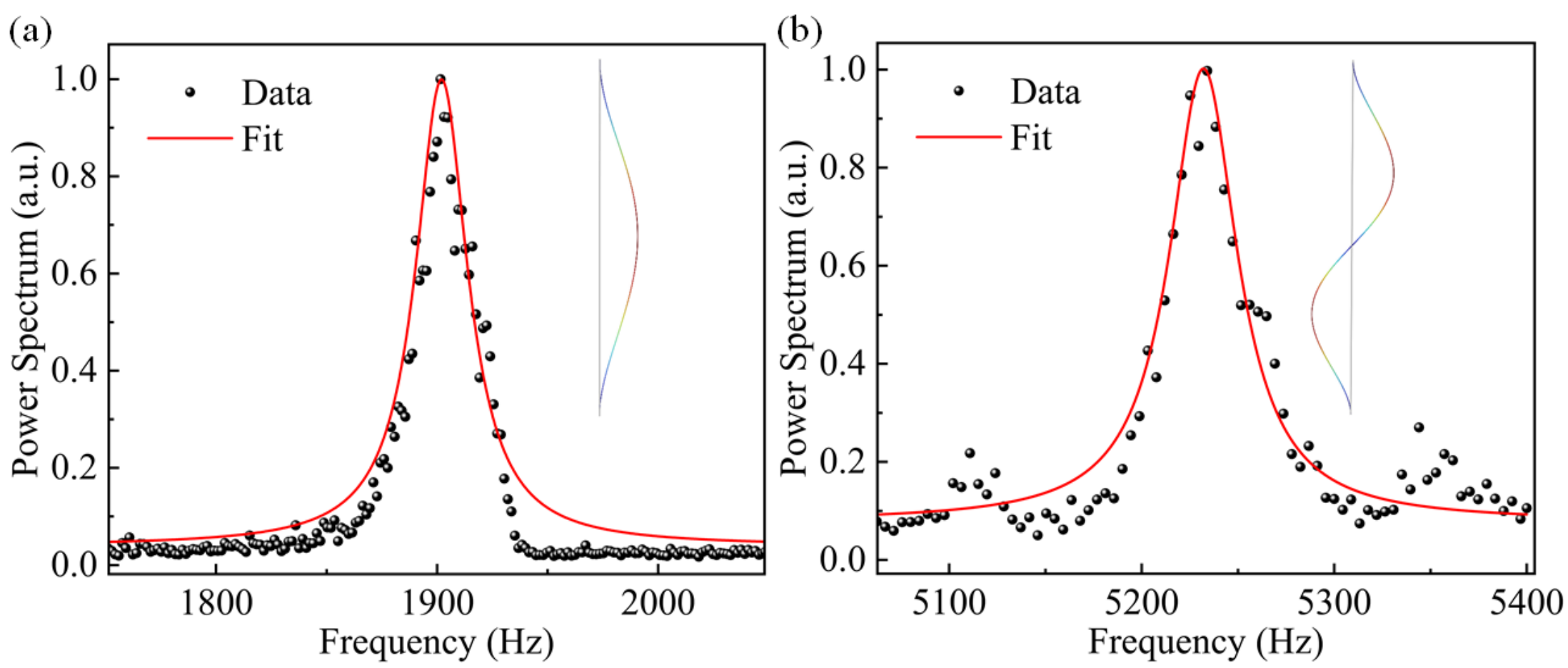

FIG. S5. Normalized mechanical power spectra of the iron-wire mechanical resonator. (a) First-order modes; (b) second-order modes. The insets show the mode shapes of the first and second modes.

Alt text:
FIG. S5: [Normalized mechanical power spectra of the iron-wire mechanical resonator. (a) First-order modes; (b) second-order modes. The insets show the mode shapes of the first and second modes.]

For the fiber mechanical resonator, we use the microfiber mechanical resonator in our previous work[2]. As shown in Fig. S6, a single-mode optical fiber is first tapered to produce a microfiber with suitable diameter, and then the microfiber is spliced with two single-mode fibers. As shown in Fig. S7, the first-order mode has a resonant frequency of 16.5 kHz and a $Q$ factor of 880,500. Meanwhile, the third-order mode has a resonant frequency of 32.9 kHz and a $Q$ factor of 601,600.

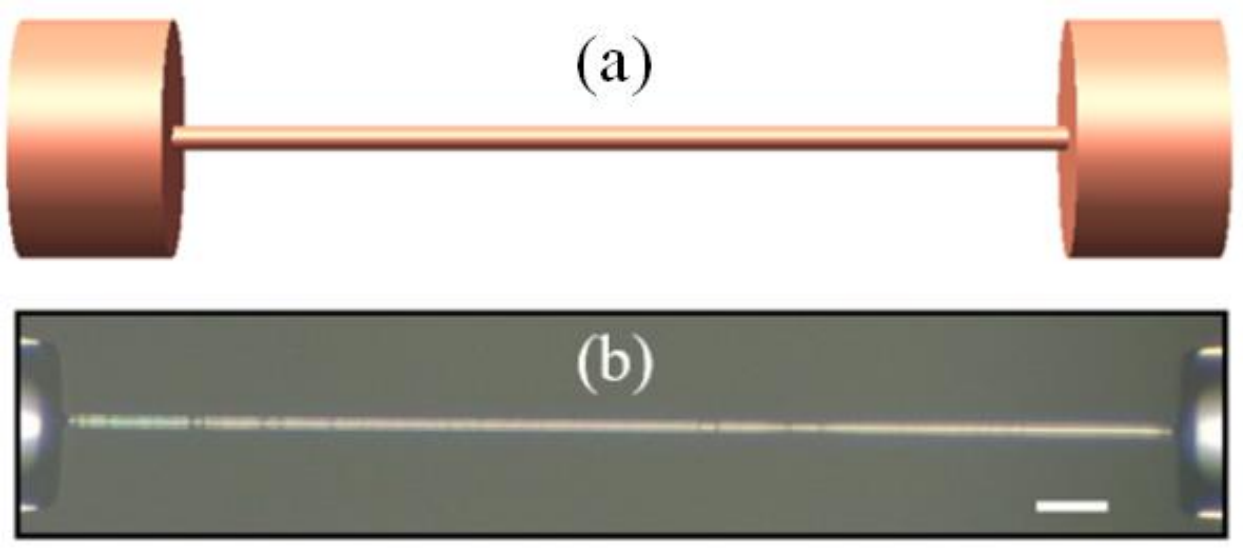

FIG. S6. (a) Schematic of the microfiber resonator. (b) Optical micrograph of the microfiber resonator. The white scale bar at the lower right corner is 50 μm.

Alt text:
FIG. S6: [Tapering a single-mode fiber to form a microfiber of suitable diameter and then splicing it with two single-mode fibers. (a) Schematic of the microfiber resonator. (b) Optical micrograph of the microfiber resonator]

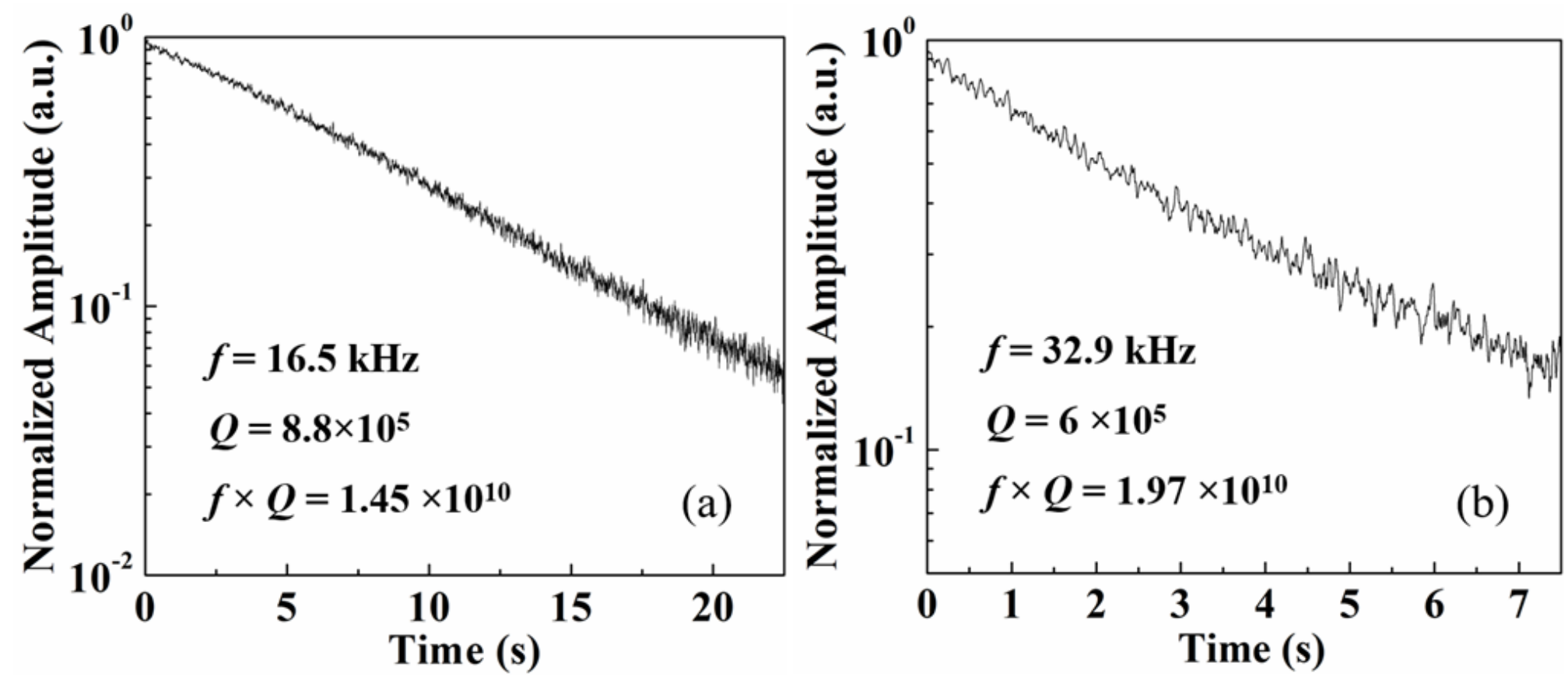

FIG. S7. Ringdown measurements for the fundamental (a) and higher-order (b) modes.

Alt text:
FIG. S7: [Ringdown measurements of the fiber mechanical resonator. (a) First-order modes; (b) third-order modes.]